# Holographic Radiance Cascades for 2D Global Illumination

Rouli Freeman
*University of Oxford*
United Kingdom
rouli.freeman@gmail.com

Alexander Sannikov
*Grinding Gear Games*
New Zealand

Adrian Margel
adrianmargel.ca

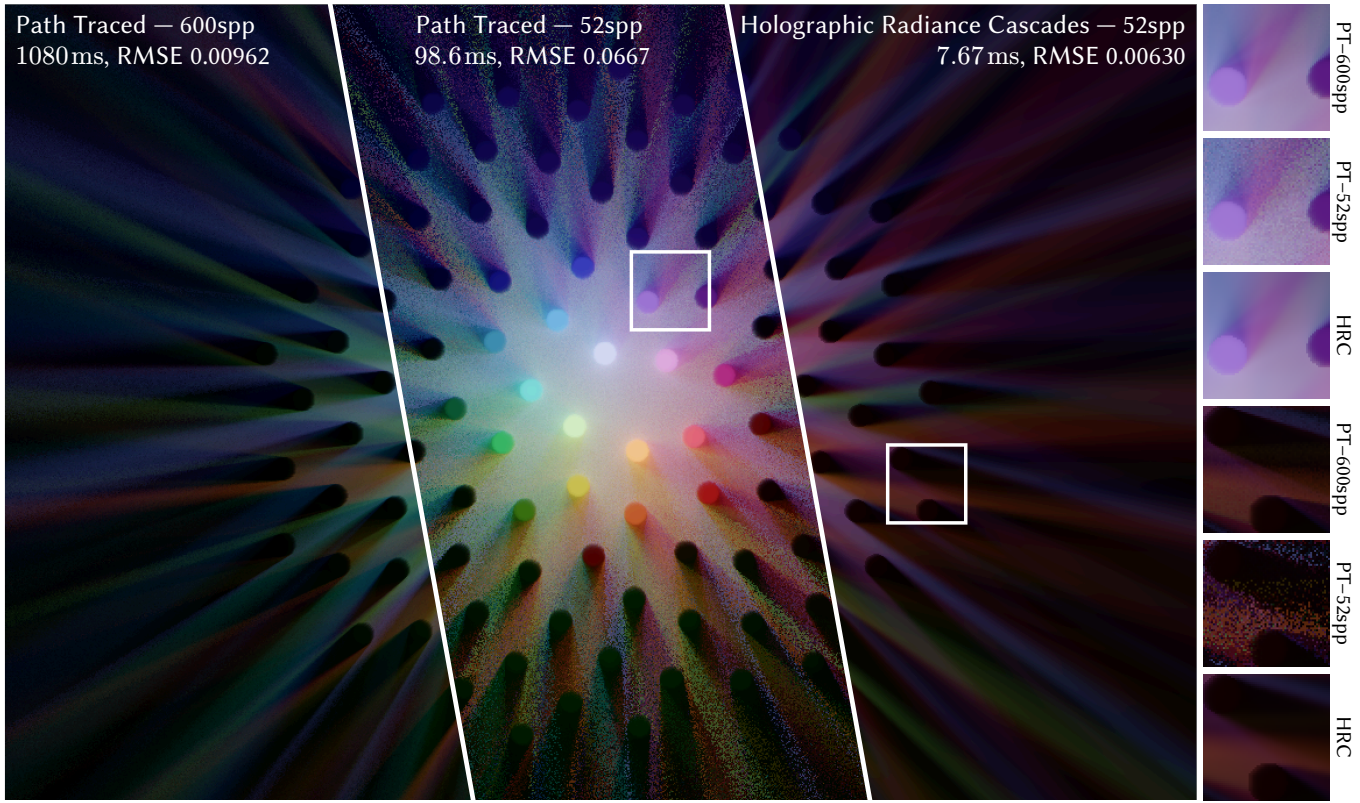

Fig. 1. A 1024 × 1024 pixel scene with multiple emissive objects, rendered using naive path tracing and Holographic Radiance Cascades (HRC). Our algorithm improves the root mean squared error (RMSE) by 10x compared to naive path tracing with an equal number of samples, where we calculate spp for HRC as the number of ray intervals calculated divided by the total number of pixels.

Efficiently calculating global illumination has always been one of the greatest challenges in computer graphics. Algorithms for approximating global illumination have always struggled to run in realtime for fully dynamic scenes, and have had to rely heavily on stochastic raytracing, spatialtemporal denoising, or undersampled representations, resulting in much lower quality of lighting compared to reference solutions. Even though the problem of calculating global illumination in 2D is significantly simpler than that of 3D, most contemporary approaches still struggle to accurately approximate 2D global illumination under realtime constraints.

We present Holographic Radiance Cascades: a new single-shot scene-agnostic radiance transfer algorithm for global illumination, which is capable of achieving results visually indistinguishable from the 2D reference solution at realtime framerates. Our method uses a multi-level radiance probe system, and computes rays via combining short ray intervals as a replacement for conventional raytracing. It runs at constant cost for a given scene size, taking 1.85 ms for a 512 × 512 pixel image and 7.67 ms for 1024 × 1024 on an RTX 3080 Laptop.

## 1 Introduction

The field of computer graphics has primarily focused on 3D Global Illumination (GI). Unfortunately, contemporary realtime approaches for 3D GI are very far from the reference solution in the general case of fully dynamic scenes without significant temporal reuse. In fact, even though the problem of 2D GI — computing the light traveling through points in a finite grid, such as in Fig. 1 — is significantly simpler than that of 3D GI, it still has no acceptable realtime single-shot solution.

We introduce a new method for this called Holographic Radiance Cascades, which can provide results nearly identical to reference in realtime for arbitrary scenes contained within the grid boundaries. It is a variant of the Radiance Cascades algorithm [Sannikov 2023], so it does not use stochastic raytracing and therefore is noiseless and does not need any temporal caching. Our algorithm improves over RC by removing redundancies in the radiance field encoding, making it more efficient and capable of handling hard shadows. We also provide a specialized acceleration structure that approximates long rays by combining short ones. This provides performance independent of scene complexity, which allows for our algorithm to handle detailed volumetrics and shapes without slowdown.

## 2 Related Work

Jarosz et al. [2012] is one of the best papers on 2D Global Illumination. It discusses how to formulate standard rendering concepts in two dimensions, and adapts path tracing, photon mapping, and irradiance caching to 2D. However, it does not concern itself with calculating fluence at non-surface points.

Radiance Cascades (RC) is a new method for global illumination introduced by Sannikov [2023], which is currently used in the game *Path of Exile 2* [Grinding Gear Games 2024] for screenspace illumination. Its approach is very well suited to 2D, as the simplest formulation of its algorithm tracks the fluence across space, rather than on surfaces. RC works somewhat similarly to irradiance probe methods [Majercik et al. 2019], although it stores values in multiple cascades with varying spatial and angular resolutions, which allows the algorithm to handle both ambient occlusion effects and large-scale environmental lighting with the same approach. In its default formulation, it has noticeable artifacts manifesting as rings around light sources, but there are ways to remove these, such as the "Bilinear Fix" described in Osborne and Sannikov [2024]. Radiance Cascades is particularly good at handling diffuse GI, as its behavior is independent of the number of light sources, however it has errors when resolving small penumbras.

## 3 Background

We seek to calculate the fluence $F$ at positions in a 2D world, defined as the radiance integrated across all directions at a point:

$$F(\boldsymbol{p}) := \int_0^{2\pi} L(\boldsymbol{p}, \vec{\omega}(\theta)) d\theta \quad (1)$$

where $L(\boldsymbol{p}, \vec{\omega})$ is the radiance at $\boldsymbol{p}$ arriving from the normalized direction vector $\boldsymbol{\omega}$, and $\vec{\omega}(\theta)$ is the direction at angle $\theta$. In comparison to irradiance, this does not have a cosine term, as there may not be a surface at $\boldsymbol{p}$. For simplicity, we will not concern ourselves with multiple wavelengths of light, as they are typically independent.

In our model, we assume the world consists entirely of participating media, as surfaces can be modelled using highly-dense volumes by taking a modified version of the BRDF as the source function. Thus, we define the transmittance $T_r$ as the fraction of radiance transmitted between two points [Pharr et al. 2023, sec 11.2]:

$$T_r(\boldsymbol{p} \leftarrow \boldsymbol{q}) := \exp\left(-\int_0^d \sigma_t(\boldsymbol{p} + t\vec{\omega}, -\vec{\omega}) dt\right) \quad (2)$$

where $d = \|\boldsymbol{q} - \boldsymbol{p}\|$, $\vec{\omega} = (\boldsymbol{q} - \boldsymbol{p})/d$ is the normalized direction vector, and $\sigma_t(\boldsymbol{p}, \vec{\omega})$ is the attenuation coefficient for the volume at position $\boldsymbol{p}$ in direction $\vec{\omega}$.

We can then compute the radiance by integrating the source function $L_s(\boldsymbol{p}, \vec{\omega})$, describing the emission and in-scattering at $\boldsymbol{p}$ in direction $\vec{\omega}$ over the ray:

$$L(\boldsymbol{p}, \vec{\omega}) = \int_0^\infty T_r(\boldsymbol{p} \leftarrow \boldsymbol{p} + t\vec{\omega}) L_s(\boldsymbol{p} + t\boldsymbol{\omega}, -\vec{\omega}) dt \quad (3)$$

Following the Radiance Cascades framework [Sannikov 2023], we define the radiance interval $L_r(\boldsymbol{p} \leftarrow \boldsymbol{q})$ as the radiance contributed by the line segment from $\boldsymbol{q}$ to $\boldsymbol{p}$ rather than an infinite ray:

$$L_r(\boldsymbol{p} \leftarrow \boldsymbol{q}) := \int_0^d T_r(\boldsymbol{p} \leftarrow \boldsymbol{p} + t\vec{\omega}) L_s(\boldsymbol{p} + t\vec{\omega}, -\vec{\omega}) dt \quad (4)$$

where $d = \|\boldsymbol{q} - \boldsymbol{p}\|$ and $\vec{\omega} = (\boldsymbol{q} - \boldsymbol{p})/d$ is the normalized direction vector. Then, $L(\boldsymbol{p}, \vec{\omega}) = \lim_{d \to \infty} L_r(\boldsymbol{p} \leftarrow \boldsymbol{p} + d\vec{\omega})$ simply by the definition of the integral (although this would not be the case if there was an environment map).

Now, let $\boldsymbol{p}$, $\boldsymbol{q}$, and $\boldsymbol{r}$ be colinear points, with $\boldsymbol{q}$ in the middle. By expanding Eq. 2 and using $\int_a^c \cdot = \int_a^b \cdot + \int_b^c \cdot$, we get that

$$T_r(\boldsymbol{p} \leftarrow \boldsymbol{r}) = T_r(\boldsymbol{p} \leftarrow \boldsymbol{q}) \cdot T_r(\boldsymbol{q} \leftarrow \boldsymbol{r}) \quad (5)$$

Using Eq. 4 and Eq. 5, we can also get a similar rule for radiance intervals:

$$L_r(\boldsymbol{p} \leftarrow \boldsymbol{r}) = L_r(\boldsymbol{p} \leftarrow \boldsymbol{q}) + T_r(\boldsymbol{p} \leftarrow \boldsymbol{q}) \cdot L_r(\boldsymbol{q} \leftarrow \boldsymbol{r}) \quad (6)$$

So, we can compute the radiance and transmittance of an interval by decomposing it into smaller intervals. We show an example of this in Fig. 2. Also notice that we can use Eq. 6 when $\boldsymbol{q}$ is arbitrarily far away, which allows us to compute $L(\boldsymbol{p}, \vec{\omega})$ using the value at $L(\boldsymbol{p} + t\vec{\omega}, \vec{\omega})$ and the radiance and transmittance of the in-between interval. Defining $\langle r, t \rangle$ to be the combined radiance interval and transmittance pair, we express these as two formulas:

$$\mathbf{Merge}(\langle r_n, t_n \rangle, \langle r_f, t_f \rangle) := \langle r_n + t_n \cdot r_f, t_n \cdot t_f \rangle \quad (7)$$

$$\mathbf{Merge}_r(\langle r_n, t_n \rangle, r_f) := r_n + t_n \cdot r_f \quad (8)$$

Notice how these are identical to the formulas for premultiplied alpha blending, where the transmittance is $1 - \alpha$ [Porter and Duff 1984].

For convenience, we will define

$$\mathbf{Trace}(\boldsymbol{p}, \boldsymbol{q}) = \langle L_r(\boldsymbol{p} \leftarrow \boldsymbol{q}), T_r(\boldsymbol{p} \leftarrow \boldsymbol{q}) \rangle \quad (9)$$

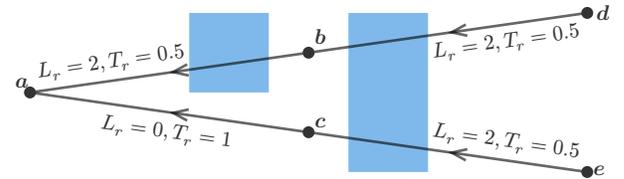

Fig. 2. In this scene, both blocks attenuate the ray by 50%, and have an emission of $L_s(\boldsymbol{p}, \vec{\omega}) \cdot \sigma_t(\boldsymbol{p}, \vec{\omega}) = 4$. Then, we have $L_r(\boldsymbol{a} \leftarrow \boldsymbol{e}) = 2$, $T_r(\boldsymbol{a} \leftarrow \boldsymbol{e}) = 0.5$, $T_r(\boldsymbol{a} \leftarrow \boldsymbol{d}) = 0.25$, and $L_r(\boldsymbol{a} \leftarrow \boldsymbol{d}) = 3$.

### 3.1 Radiance Cascades

Assume that all objects in the world have a minimum size of 1, and that they are within $a$ distance from the origin. In order to accurately approximate the fluence at the origin, we must use at least $2\pi a$ samples at the furthest edge to ensure that no features are missed. The simplest way of doing this is by dividing the world into $I = \lceil 2\pi a \rceil$ cones:

$$F(\boldsymbol{0}) = \sum_{i=0}^{I-1} F_i \quad (10)$$

where $F_i$ is the angular fluence for the $i$th cone, being the radiance integrated from $2\pi(i-1)/I$ to $2\pi i/I$. We approximate it by sam-pling the radiance in the center of the cone (where $\vec{\omega}(\theta)$ is the direction at angle $\theta$):

$$F_i = \frac{2\pi}{I} L\left(\mathbf{0}, \vec{\omega}\left(\frac{2\pi(i + 0.5)}{I}\right)\right) \quad (11)$$

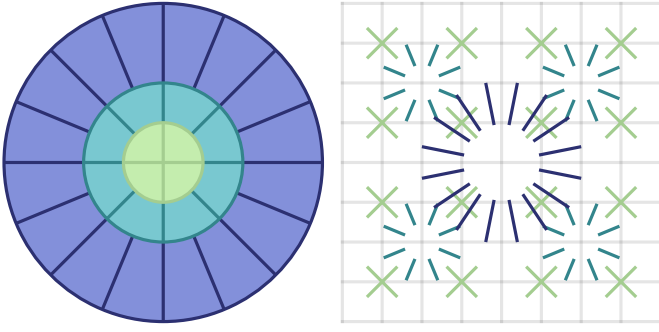

Fig. 3. The conical frustums used to approximate the fluence at a point in Radiance Cascades (left), and the rays used for three separate layers of probes (right) — shrunken for visibility. Note that each cascade has a quarter the probes of the previous level.

However this is inefficient, as at a distance of $\frac{a}{2}$, there only needs to be $\pi a$ samples to resolve the scene. Instead, assuming that $I$ is even, we may separate the world into $J = \frac{I}{2}$ cones, each of which is then further subdivided into 2 conical frustums after passing a distance of $\frac{a}{2}$, and approximate all of these with single ray casts. Consider the $j$th cone, with central axis $\boldsymbol{\omega}$, and let $F_+$ and $F_-$ be the angular fluence of the frustums, approximated by tracing the centers of them starting from $\frac{a}{2}$. We then approximate the angular fluence and transmittance of the cone from 0 to $\frac{a}{2}$ as $\langle f_n, t_n \rangle = \frac{2\pi}{I} \times \mathbf{Trace}(\mathbf{0}, \frac{a}{2}\vec{\omega})$ where $A \times \langle r, t \rangle = \langle Ar, t \rangle$ converts the radiance into angular fluence with arc $A$ and leaves the transmittance unchanged, and $\mathbf{Trace}$ is defined in Eq. 9.

The total angular fluence is then calculated as

$$F_j = \mathbf{Merge}_r(\langle f_n, t_n \rangle, F_+ + F_-) \quad (12)$$

as $\mathbf{Merge}$ can also operate on fluence values if they have the same angular size.

We may then recursively repeat this for the 0 to $\frac{a}{2}$ cone, stopping when the cone length is less than the feature resolution, as shown in Fig. 3 (left).

In standard Radiance Cascades (RC) [Sannikov 2023; Osborne and Sannikov 2024], we discretize the world into multiple cascades of grids of probes, each of which stores the fluence accumulated within an exponentially-increasing interval (with the exception of the base grid, which starts at 0), with an angular discretization proportional to the starting distance, as in Fig. 3. We compute each cascade from the next higher cascade, by tracing from the starting point and then merging as in Eq. 12, with bilinear interpolation used when probes are at different positions. The most common setting for RC in 2 dimensions is to have each cascade level's probe grid have half the spatial resolution of the previous level — using the observation that under most conditions, light from far away sources is slowly varying — and quadruple the angular resolution (which means that to make the ray spacing uniform, the starting distance also increases in powers of 4). For a fixed area of interest, this makes each cascade store the same number of values. We then pick the number of cascades such that the combined ray length exceeds the area of interest size.

However, standard RC cannot handle distant lights with a low angular resolution. Let the base probe spacing be $A$ and ray length be $B$, and then consider a light of a small size $Y$ at a distance of $X \cdot B$, occluded by an object at distance $\frac{X \cdot B}{2}$ such that the origin is in the penumbra, as shown in Fig. 5. This penumbra will have width $Y$ near the origin, however the probes that resolve the occluder will be in cascade $\approx \log_4(X)$, and as such have spacing $\sqrt{X} \cdot A$, which means that if $\sqrt{X} \cdot A > Y$, RC is incapable of resolving it accurately and produces interpolation artifacts, as can be seen in Fig. 8.

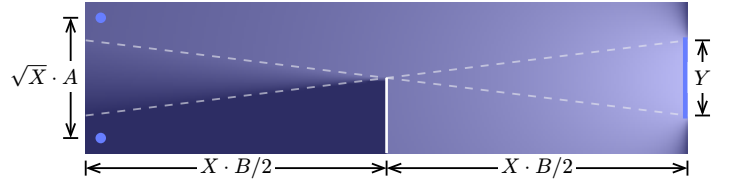

Fig. 5. The penumbra of a small light source. It's impossible to reconstruct the penumbra accurately near the left side by interpolating values at these probes when $\sqrt{X} > Y$.

## 4 Methodology

In Holographic Radiance Cascades (HRC), we solve this problem by adjusting the probe positions so that there is always a high spatial resolution of probes perpendicular to the angles the probe is gathering from — we do not need to do this in both directions, as a penumbra that is varying in both axes fast must be from a nearby light source, and so is picked up by a lower cascade.

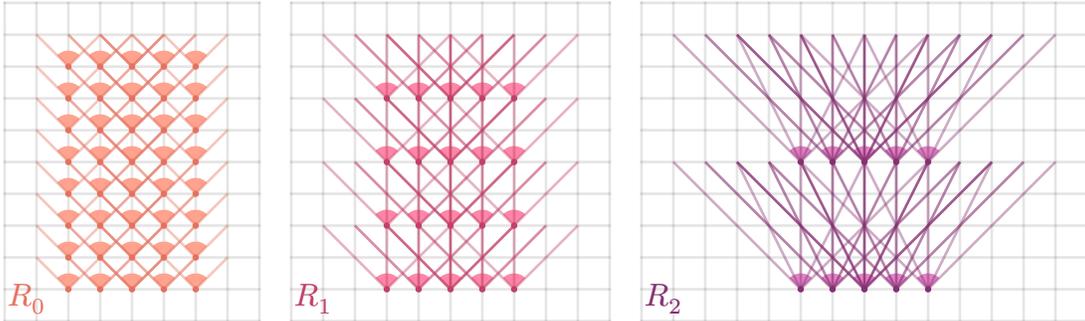

Fig. 4. The spatial positioning and extent of the first three cascades of probes in HRC, where the $x$-axis is vertical.

To do this, we split the fluence arriving at the base layer of probes into 4 quadrants, which can be combined at the end. Without loss of generality, assume that we are only considering the angular fluence between $-\frac{\pi}{4}$ and $\frac{\pi}{4}$. In contrast to normal RC, which reduces the spatial resolution in both directions every cascade level, we only reduce the spatial resolution in the direction parallel to the probe's facing — so the $n$th level has probes at positions $\boldsymbol{p} = (x \cdot 2^n, y)$, for integer $x, y$. We then increase the angular resolution by a factor of 2 every level, as well as the distance each level traces before merging, resulting in the same spacing between ray ends at every level.

More specifically, for the $n$th cascade, let $i$ be the index of the direction taking on half-integer values between $\frac{1}{2}$ and $2^n - \frac{1}{2}$, inclusive. Then, let $R_n(\boldsymbol{p}, i)$ be our approximation to the angular fluence of the cone starting at the probe $\boldsymbol{p}$, with edges passing through $\boldsymbol{p} + \vec{v}_n(i + \frac{1}{2})$ and $\boldsymbol{p} + \vec{v}_n(i - \frac{1}{2})$, as shown in Fig. 4, where $\vec{v}_n(k) := (2^n, 2k - 2^n)$ is the offset of the next probe in direction $k$. For a given probe, the combination of all of its cones then spans the entire arc between $-\frac{\pi}{4}$ and $\frac{\pi}{4}$. If we define $\mathrm{angle}([x, y]) := \tan^{-1}(y/x)$ as the angle of a point from the origin, we may compute the angular size of the cone in the $i$th direction of a probe at cascade $n$ using this formula:

$$A_n(i) := \mathrm{angle}\left(\vec{v}_n\left(i + \frac{1}{2}\right)\right) - \mathrm{angle}\left(\vec{v}_n\left(i - \frac{1}{2}\right)\right) \quad (13)$$

To compute our approximation to the angular fluence, we define $R_n$ recursively in terms of $R_{n+1}$; conceptually, we do this by splitting the cone through the middle, and then approximating each side by a thinner cone in the next cascade level: For a fixed $\boldsymbol{p} = (x \cdot 2^n, y)$ and $i$, let $R_n(\boldsymbol{p}, i) = F_+ + F_-$, where $F_+$ and $F_-$ are the contributions from the upper and lower halves of the cone respectively. This has to be handled differently for odd and even $x$.

For odd $x$, there is no information for the $n + 1$th cascade at $\boldsymbol{p}$, so we compute the cones by tracing along the edges of the cone until we hit a probe in the higher cascade, which happens at $\boldsymbol{q}_\pm = \boldsymbol{p} + \vec{v}_n(i \pm \frac{1}{2})$ (with an x-coordinate of $(x + 1) \cdot 2^n$) and then merging it with the angular fluence at that point in the closest direction, which has index $j_\pm = 2i \pm \frac{1}{2}$, as shown in Fig. 6:

$$F_\pm = \mathbf{Merge}_r(A_{n+1}(j_\pm) \times \mathbf{Trace}(\boldsymbol{p}, \boldsymbol{q}_\pm), R_{n+1}(\boldsymbol{q}_\pm, j_\pm)) \quad (14)$$

For even $x$, it would be obvious to simply use $F_\pm = R_{n+1}(\boldsymbol{p}, j_\pm)$, since those perfectly match up with the upper and lower halves of the $R_n(\boldsymbol{p}, i)$ cone. Unfortunately, this causes artifacts: the value of $R_n$ computed using the previous formula for odd $x$ has a bias towards the edges of the cone, so simply combining two cones would cause a bias towards the middle that cannot be corrected later. Instead, since $R_{n+1}$ does not exist at $(x + 1) \cdot 2^n$, we generate the fluence by interpolating along the line from $\boldsymbol{p}$ to $\boldsymbol{q}_\pm = \boldsymbol{p} + 2\vec{v}_n(i \pm \frac{1}{2})$ as follows:

$$F_\pm = (F_\pm^0 + F_\pm^1)/2 \quad (15)$$

where:

$$F_\pm^0 := R_{n+1}(\boldsymbol{p}, j_\pm) \quad (16)$$

$$F_\pm^1 := \mathbf{Merge}_r(A_{n+1}(j_\pm) \times \mathbf{Trace}(\boldsymbol{p}, \boldsymbol{q}_\pm), R_{n+1}(\boldsymbol{q}_\pm, j_\pm)) \quad (17)$$

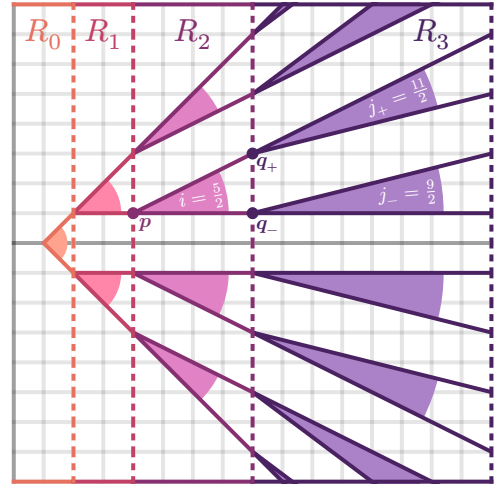

Fig. 6. The first 4 cascades of $R$ used to calculate $R_0([0, 1], 0)$, showing in particular the values used to compute $R_2(\boldsymbol{p} = [4, 1], \frac{5}{2})$ using Eq. 14.

### 4.1 Acceleration Structure

It is also possible to approximate the rays traced to generate the HRC angular fluence values in an efficient manner:

For integers $n$, $x$, $y$, and $k \leq 2^n$, let $T_n(\boldsymbol{p}, k)$ store an approximation of $\mathbf{Trace}(\boldsymbol{p}, \boldsymbol{p} + \vec{v}_n(k))$, where $\boldsymbol{p} = (x \cdot 2^n, y)$. Then, for odd $x$, we can replace the use of $\mathbf{Trace}(\boldsymbol{p}, \boldsymbol{p} + \vec{v}_n(i \pm \frac{1}{2}))$ in $F_\pm$ with $T_n(\boldsymbol{p}, i \pm \frac{1}{2})$. Similarly, for even $x$, $\mathbf{Trace}(\boldsymbol{p}, \boldsymbol{p} + 2\vec{v}_n(i \pm \frac{1}{2}))$ can be replaced with $T_{n+1}(\boldsymbol{p}, 2i \pm 1)$, since $2\vec{v}_n(k) = \vec{v}_{n+1}(2k)$.

Now, we define $T_{n+1}$ recursively in terms of $T_n$ similarly to our definition of $R_n$, as shown in Fig. 7. If $2k$ is even, we can calculate $T_{n+1}(\boldsymbol{p}, 2k)$ without any further approximation:

$$T_{n+1}(\boldsymbol{p}, 2k) = \mathbf{Merge}(T_n(\boldsymbol{p}, k), T_n(\boldsymbol{p} + \vec{v}_n(k), k)) \quad (18)$$

as we can represent any ray as a combination of the near and far halves. If $2k$ is odd, then the direction $k$ is not an integer, so we cannot perform this. Instead, we blend the two closest approximations: Let

$$F_\pm := \mathbf{Merge}(T_n(\boldsymbol{p}, k \pm \tfrac{1}{2}), \\ T_n(\boldsymbol{p} + \vec{v}_n(k \pm \tfrac{1}{2}), k \mp \tfrac{1}{2})) \quad (19)$$

and then we can compute $T_{n+1}$ as:

$$T_{n+1}(\boldsymbol{p}, 2k) = (F_- + F_+)/2 \quad (20)$$

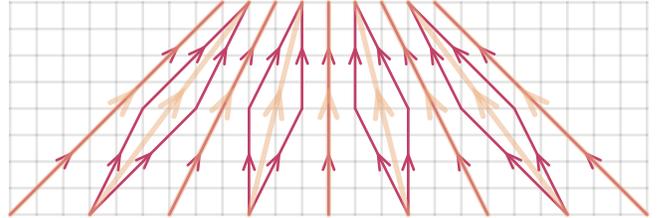

Fig. 7. Combining rays from $T_2$ to generate $T_3$, where the $x$-axis is vertical.

### 4.2 Implementation

**Algorithm 1:** Single bounce lighting using HRC

**Procedure** HRC($X, Y$):
  **for** $n$ **in** 0 **to** 2:
    **for** $x$ **in** $0..\lceil X/2^n \rceil$, $y$ **in** $0..Y$, $k$ **in** $0..2^n + 1$:
      let $p = (x \cdot 2^n, y)$
      $T_n(p, k) := \mathbf{Trace}(p, p + \vec{v}_n(k))$
  **for** $n$ **in** 3 **to** $\lceil \log_2(X) \rceil$:
    **for** $x$ **in** $0..\lceil X/2^n \rceil$, $y$ **in** $0..Y$, $k$ **in** $0..2^n + 1$:
      let $p = (x \cdot 2^n, y)$
      **if** $k$ **is even**:
        Compute $T_n(p, k)$ from $T_{n-1}$ using Eq. 18
      **else**:
        Compute $T_n(p, k)$ from $T_{n-1}$ using Eq. 20
  **for** $n$ **in** $\lceil \log_2(X) \rceil - 1$ **to** 0:
    **for** $x$ **in** $0..\lceil X/2^n \rceil$, $y$ **in** $0..Y$, $i$ **in** $0..2^n$:
      let $p = (x \cdot 2^n, y)$
      **if** $x$ **is even**:
        Compute $F_+, F_-$ from $R_{n+1}, T_{n+1}$ using Eq. 15
      **else**:
        Compute $F_+, F_-$ from $R_{n+1}, T_n$ using Eq. 14
      $R_n(p, i) = F_+ + F_-$
  **return** $R_0$

**Procedure** Lighting($X, Y$):
  **for** dir **in** 0..4:
    $R_0 := \mathbf{HRC}(X, Y)$
    **for** $x$ **in** $0..X$, $y$ **in** $0..Y$:
      $L([x, y]) \mathrel{+}= R_0([x+1, y], 0)$
    Rotate world by 90°.
    Rotate $L$ by 90°.
    Swap $X, Y$.
  **return** $L$

Assume that we want to compute fluence for a $X \times Y$ grid, where all light sources are positioned within it. In order to do this, we split the fluence into the four quadrants, which we handle separately. Consider the angular fluence arriving from the +x quadrant. For an integer position $[x, y]$, we can approximate this using the first level cascade information, $R_0([x+1, y], 0)$ — this offset is necessary to prevent bias as can be seen in Fig. 12: if it was removed, the diagonal rays of different quadrants would overlap, which results in crosses of increased brightness. $R_0$ then depends on values $R_1...R_{N-1}$ where $N = \lceil \log_2(X) \rceil$ — any lookups of $R_N$ or greater would be at $x = 2^N$, and so are uniformly 0 as there are no lights beyond the region.[1] In order to compute these, we approximate the $\mathbf{Trace}(p, p + \vec{v}_n(k))$ values using the acceleration structure described in Section 4.1. We manually initialize $T_n(p, k) = \mathbf{Trace}(p, p + \vec{v}_n(k))$ using a standard raytracing method for $n = 0, 1, 2$ to reduce error, as the cost to trace the short rays is relatively small and the error from odd-direction merging is magnified the shorter the ray is.[2] For $n = 3...N$, we compute $T_n$ from $T_{n-1}$ using Eq. 18 and Eq. 20 — note that we need $T_N$ in order to evaluate even $x$ values of $R_{N-1}$. Then, from $n = N - 1$ to $n = 0$, we evaluate $R_n$ using $R_{n+1}, T_n$, and $T_{n+1}$, using Eq. 14 and Eq. 15 (treating $R_N$ as uniformly 0). Finally, after summing up all 4 quadrants of fluence, we apply a 1px cross blur, using the kernel

$$\frac{1}{8}\begin{bmatrix} 0 & 1 & 0 \\ 1 & 4 & 1 \\ 0 & 1 & 0 \end{bmatrix} \qquad (21)$$

and ignoring the probes that differ significantly in opacity from the target. This is necessary, as Holographic Radiance Cascades results in checkerboard artifacts — notice that $\vec{v}_n(k)$ is always a multiple of 2 for any $n \geq 1$, which results in the probes with odd and even $y$ values not interacting; for example, in Fig. 6, all probes except for the first level have odd $y$ values.

For multiple diffuse bounces, we can simply feed the output fluence values into another iteration of the algorithm (which can be done temporally). However, the output fluence only has 4 directions, which prevents it from being used for specular reflections. This could be solved by looking up the higher cascade angular fluence values instead, although we have not implemented that.

### 4.3 Performance Analysis

We will compute the theoretical time and space for this implementation. For a $X \times X$ size grid, where $X = 2^N$ for an integer $N$, $R_n$ has $2^{N-n} \cdot 2^N \cdot 2^n = 4^N = X^2$ values, and $T_n$ has $2^{N-n} \cdot 2^N \cdot (2^n + 1) = 4^N \cdot (1 + 2^{-n}) \leq 2X^2$ values, and there are $N$ total cascades of $R$, and $N + 1$ cascades of $T$.

*4.3.1 Speed.* Each cascade requires executing a constant-time algorithm per value within $R_n$ to compute it from $R_{n+1}$, which requires $O(N \cdot X^2)$ total time. This also depends on computing $T_n$, which requires $O((N + 1) \cdot 2X^2)$ time as well, resulting in $O(X^2 \log X)$ total time. Computing $T_0$ to $T_2$ also requires $4 \cdot X^2 \cdot \left(3 + 1 + \frac{1}{2} + \frac{1}{4}\right) = 19X^2$ invocations of an unaccelerated **Trace** (multiplying by the number of quadrants), for ray lengths no longer than $4\sqrt{2}$, with the majority of them being at most half as long.

*4.3.2 Memory.* Let $V_r$ and $V_t$ be the size in bytes of a single fluence or radiance value and a single transmittance value, respectively. Then, the size of $R_n$ is $4^N \cdot V_r$, and the size of $T_n$ is $4^N \cdot (1 + 2^{-n}) \cdot (V_r + V_t)$. We only need to store 2 layers of $R$ at a single time, since our output only needs to be $R_0$, and each layer only depends on the next higher one, so the storage for the fluence is $2 \cdot 4^N \cdot V_r$. However, each layer of $R_n$ depends on $T_n$ and $T_{n+1}$, so we must store each layer of the acceleration structure separately, resulting in total storage for it being $\sum_0^N 4^N \cdot (1 + 2^{-n}) \cdot (V_r + V_t) = (N + 3 + 2^{-N}) \cdot 4^N \cdot (V_r + V_t) \approx N \cdot 4^N \cdot (V_r + V_t)$ for large $N$. This makes up the majority of the memory cost, but can be avoided via use of a different acceleration structure such as a BVH for **Trace**, which would bring it down to $2 \cdot 4^N \cdot V_r$ (discounting the cost for that acceleration structure). Note that each of the four quadrants can reuse the same memory if the grid is square.

*4.3.3 Ray Count.* Each element of $T_n$ is a ray interval used to compute the result, so we need a total of $4(N + 3 + 2^{-N})$ ray calculations per output fluence value, assuming no use of the acceleration structure. This works out to 52 per base probe for a $1024 \times 1024$ sized grid. With our acceleration structure settings,

---

[1]Handling offscreen lights can then be done by simply computing $R_N$ via a different method such as cone tracing. For environment map use (assuming no occluders outside of bounds), simply look up the value of the map in the direction of $\vec{v}_N(i)$.

[2]Importantly, the total number of directions of outgoing radiance from an object is limited by the number of rays traced (in contrast to the values generated via merging), which in this case is $4 \cdot (2^2 + 1) = 20$. This can result in degraded quality of specular reflections as the BRDF is functionally blurred.

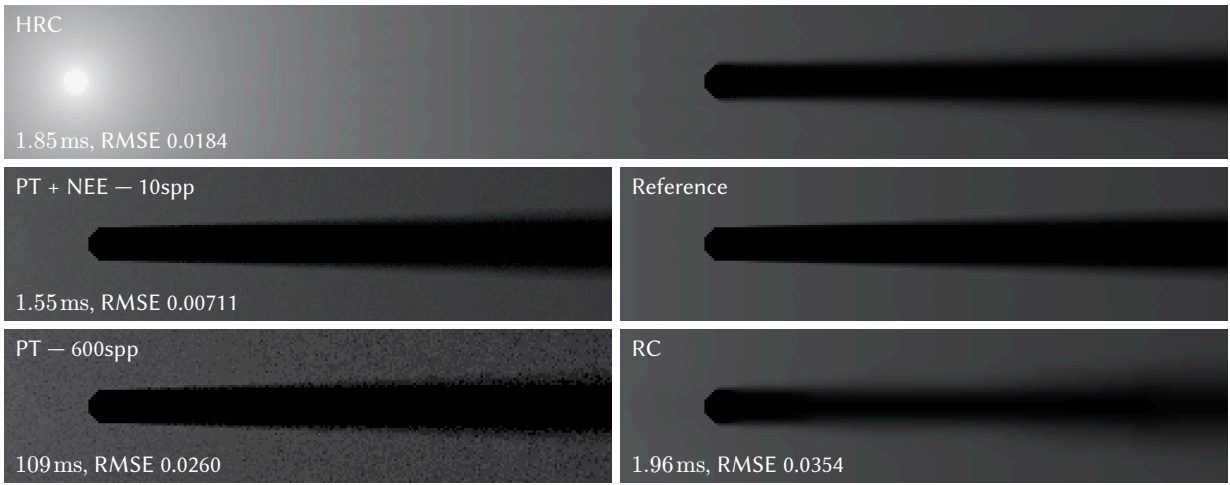

Fig. 8. The penumbra of a 14 pixel wide occluder illuminated using a 10 pixel circular light, shown using HRC, path tracing (PT), and standard RC. All results simulated on a $512 \times 512$ grid, with RMSE computed on the displayed area to accentuate differences in the penumbra.

we only need 19 rays per base probe, which are all less than 6 times the base spacing.

## 5 Results

We implement our algorithm in Rust, using the LuisaCompute framework [Zheng et al. 2022] as an abstraction over CUDA, and execute on a NVIDIA RTX 3080 Laptop GPU. We take in the scene as a $512 \times 512$ or $1024 \times 1024$ pixel image, and produce an equal-resolution fluence grid. To generate $T_0..T_2$, we trace the rays using the DDA algorithm by Amanatides and Woo [1987], and integrate the radiance and transmittance analytically within each pixel of the scene, treating it as a square of uniform extinction and emission [Pharr et al. 2023, sec 11.2]. We store the radiance and transmittance values using 3-vectors of 16-bit floats without any compression. As our algorithm does not branch based on the scene data, it has constant time for a given size, as shown in Table 1.

Note that in an actual renderer, the output fluence would likely be interpolated from these probes onto a screen 2x or 4x the size (merging with rays traced to the probes to avoid light leaks); this makes the algorithm significantly more feasible to run in combination with other GPU usage. However, we have decided to avoid this, as it detracts from the actual results.

As there are no standard test scenes for 2D, we choose to make our own. In contrast to guided path tracing algorithms, HRC functions independently from the quantity of light sources or the complexity of the geometry, so, we've chosen to test simple scenes, as that makes it easier to understand the effectiveness of our approach. Similarly, we also do not run multiple iterations for multibounce GI, apart from in Fig. 14, as that generally makes things blurrier and harder to compare.

Table 1. Performance timings for HRC on a 3080 Laptop GPU. "Merge Up" represents the computation of $T_0..T_N$, while "Merge Down" represents computing $R_{N-1}..R_0$

| Size | Merge Up | Merge Down | Total |
| --- | --- | --- | --- |
| $256 \times 256$ | 0.30 ms | 0.25 ms | 0.55 ms |
| $512 \times 512$ | 1.00 ms | 0.85 ms | 1.85 ms |
| $1024 \times 1024$ | 4.10 ms | 3.57 ms | 7.67 ms |
| $2048 \times 2048$ | 18.1 ms | 15.8 ms | 33.9 ms |

For comparison, we implement a path tracer, which uses the golden ratio low-discrepancy sequence for generating rays [Wolfe 2020], and two-level DDA with $8 \times 8$ pixel blocks to trace them [Museth 2014]. We also show use of next event estimation (NEE) [Pharr et al. 2023, sec 13.2], which we perform by sampling rays within the angle subtended by the bounding box of the light. Note that this requires knowledge about the scene, which usually would require additional computation to produce.

*5.0.1 Occluder.* Our first scene is a single light source and occluder, shown in Fig. 8. In comparison to the reference, HRC blurs the shadow by around 4 pixels, but produces nearly identical results otherwise.

We show two different path-traced images. This is the best scene for a NEE-based tracer due to the simplicity, and indeed it has the lowest error, however it still has noticeable noise in the penumbra. The naive path tracer takes a long time to converge, as the light source has a small angular size, and correspondingly has a large error. However, since it isn't guided, increasing the complexity of the scene would not reduce the quality of the output.

We also compare to an optimized Radiance Cascades implementation [Sannikov 2023], which we have adjusted for an equal-time comparison. The background gradient resolves accurately, as RC uses a similar method to ensure full coverage of the scene. However, the shadow quality is significantly worse, resulting in twice the RMSE of HRC — towards the right of the scene, the cascade level that picks up the light is increased, which results in not enough spatial resolution to pick up the penumbra.

*5.0.2 Pinhole.* In Fig. 9, we test a pinhole camera scene. For this, we compare to a path traced image using an equal number of samples per pixel; while it produces a distinguishable image, it still has plenty of noise — the majority of the light source is occluded, which results in the next event estimation producing poor results.

In contrast, HRC resolves this without any noise in a quarter of the time, which shows that it is capable of efficiently simulating large light sources and light transfer through small gaps.

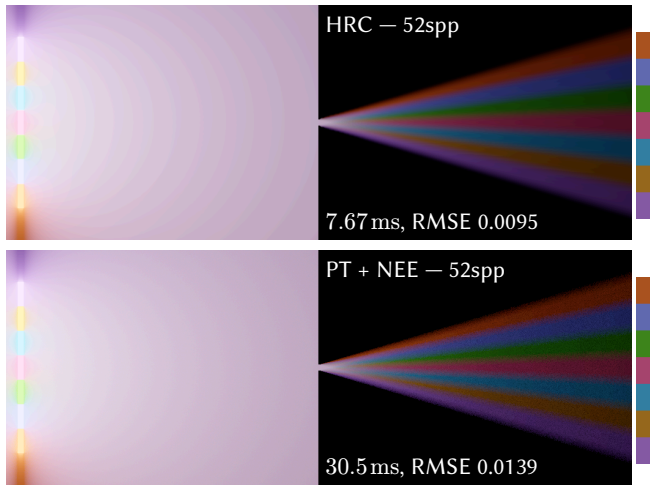

Fig. 9. Light passing through a 10 pixel wide pinhole, simulated using HRC and path tracing with next event estimation (PT + NEE), both on a 1024 × 1024 grid.

*5.0.3 Artifacts.* Our algorithm demonstrates two forms of artifacts apart from over-blurring. The most noticeable one is that it has checkerboard patterns. For example, using the scene in Fig. 8, we can see a pattern near the edge of the shadow, as in Fig. 10 (a). This is significantly lessened after applying the cross blur in Eq. 21, which results in Fig. 10 (b).

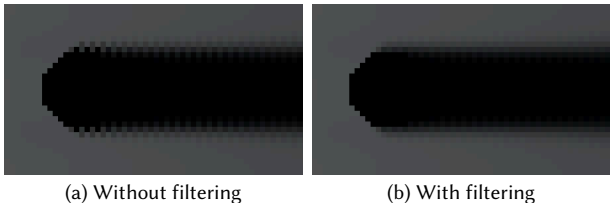

(a) Without filtering          (b) With filtering

Fig. 10. View of the occluder in Fig. 8, showing the checkerboard pattern.

It also demonstrates Moiré pattern-like behavior, especially with small lights. This is due to aliasing from the ray segments being fixed — a small change in the displayed pixel could result in a significant change in brightness, as the light could change from intersecting 2 rays to 3, for example. This can be seen in Fig. 11. Luckily, this is much less of an issue for large light sources as they have more samples, and generally ceases to be a concern for sources larger than 8 times the base probe resolution: see how Fig. 1 does not have this, and in Fig. 8, it is barely visible.

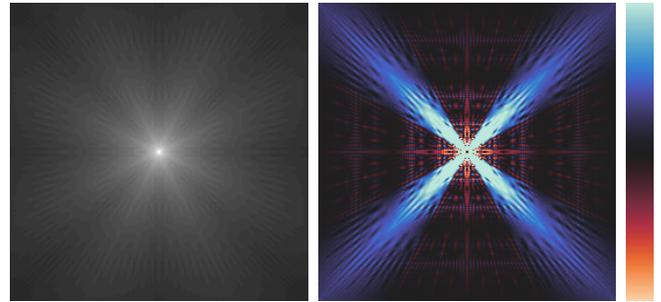

Fig. 11. A 2x2-pixel light, rendered using HRC (left), as well as the difference between it and a reference image, with the scale measuring from $-0.2\%$ to $0.2\%$ of the light intensity.

*5.0.4 Volumetrics.* A major benefit of our algorithm is that it can handle detailed volumetrics efficiently. Common acceleration structures rely on empty space skipping, which does not work for volumes that continuously vary in material, such as clouds. For example, the scene in Fig. 13 has high detail across the entire domain, due to the varying levels of opacity and the fractal surface. This results in path tracing taking 12x as long for the same number of samples (1310 ms for 600spp), compared to a mostly-empty scene such as Fig. 8, while our algorithm is unchanged.

# 6 Conclusion

We have demonstrated a new method for 2D Global Illumination that provides consistent high quality for all scenes with a large enough feature size. We do this by reformulating Radiance Cascades using a grid that decreases in resolution in only one direction to avoid shadow artifacts, and provide an acceleration structure that does not rely upon empty space skipping, which allows for consistent performance and simple volumetric handling.

The primary limitation of the current formulation of HRC is the artifacts that become visible when light sources are smaller than 8 times the base probe resolution. It also has poor scaling in 3 dimensions, taking up $N^4$ memory for a $N \times N \times N$ scene — although it remains feasible for small volumes, as seen in Fig. 15. For future work, it is worth investigating methods of distributing the computation across frames, such as caching unchanged parts of the world, or jittering rays in order to increase accuracy. The acceleration structure currently takes up almost all the memory footprint, so compressing it, or using a different method would allow for larger scenes. Finally, combining HRC with a glossy GI method would significantly improve sharp reflections.

## Acknowledgements

We would like to thank Christopher Osborne, Asbjørn Lystrup, and Manu Udupa for their assistance in proofreading.

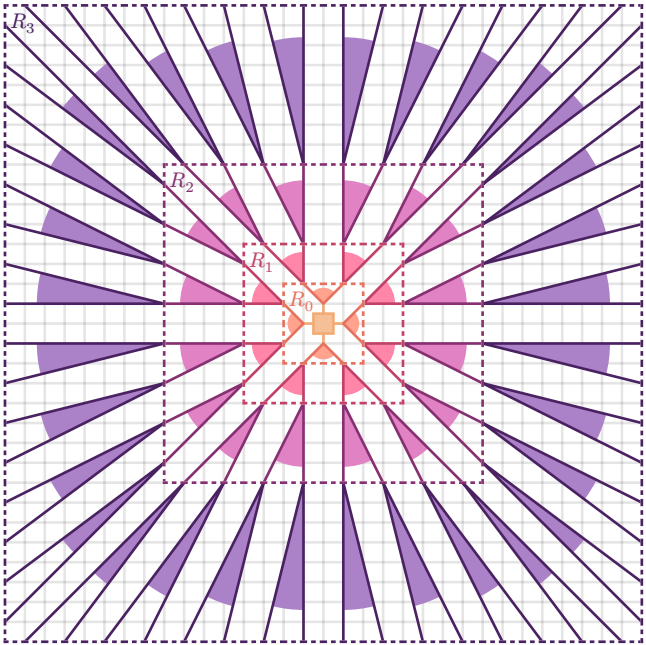

Fig. 12. The cones that make up the fluence at a single pixel computed using HRC, using the merging strategy for odd $x$ — the equivalent for RC is shown in Fig. 3 (left), although interpolation would be used at every level. Note how the traced rays (the cone edges) cover the entire area semi-uniformly.

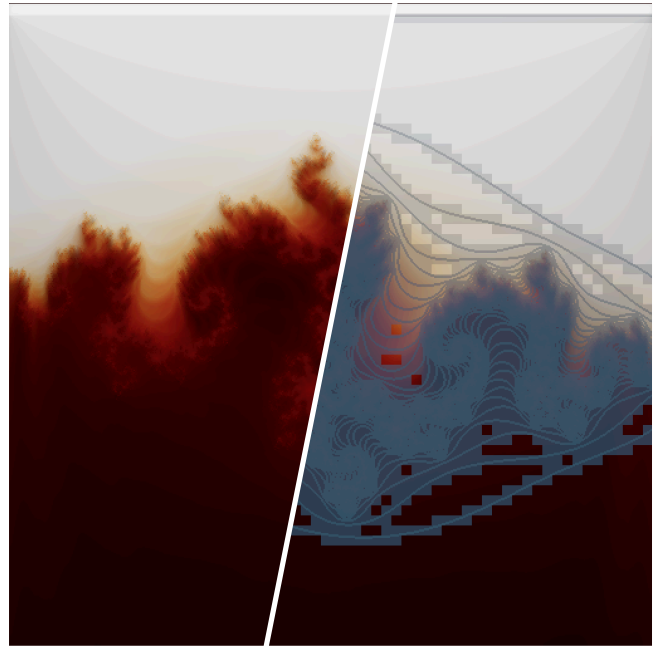

Fig. 13. A rendering of the Julia set for $c = -0.835 - 0.2321i$ with volumetric parameters determined by the escape iteration of the pixel (left), and the occupied blocks in the 2-level DDA acceleration structure (right). HRC produces results indistinguishable from the reference (RMSE 0.00498). Simulated on a $512 \times 512$ grid.

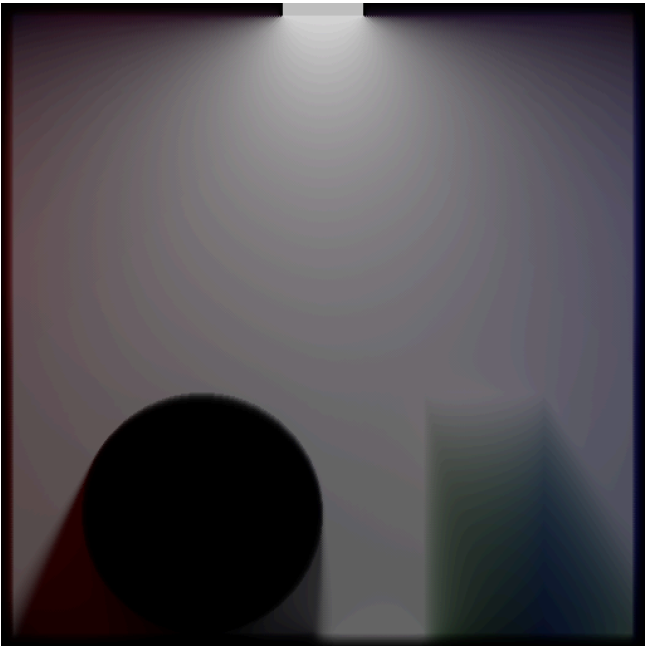

Fig. 14. A $512 \times 512$ cornell box with a solid circle and a volumetric scattering rectangle, with diffuse reflecting walls. The shadows are not perfectly dark due to the reflections off of the sides. Converged after 6 bounces (temporally accumulated).

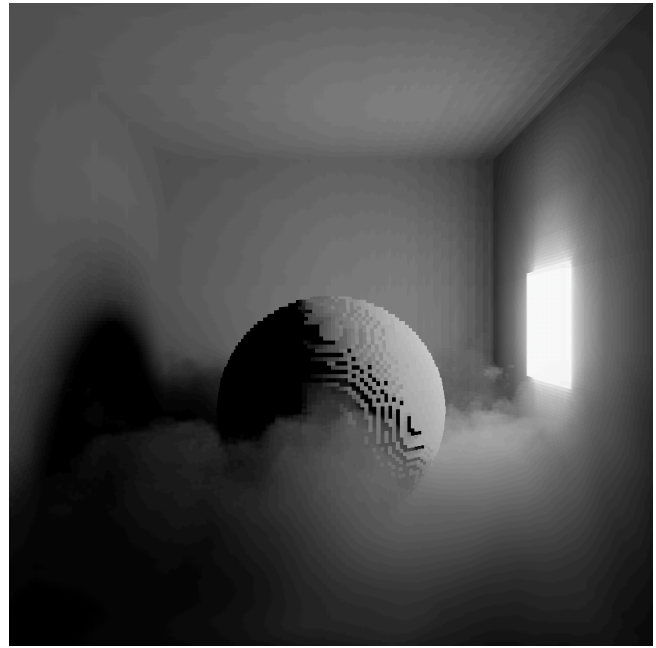

Fig. 15. A single-bounce rendering of a $128 \times 128 \times 128$ voxel scene using an experimental 3D HRC implementation. Note that the volumetrics can be done without any added cost. Surface normals have not been implemented, which results in the irregular shadowing.